\documentclass{aa}
\begin{document}

   \thesaurus{13    
              (02.19.1;  
               13.07.1)} 
   \title{Gamma Ray Bursts {\it vs.} Afterglows}

   \author{J. I. Katz}

   \offprints{J. I. Katz}

   \institute{Department of Physics and McDonnell Center for the Space
	      Sciences, Washington University, St. Louis, Mo. 63130\\
              email: katz@wuphys.wustl.edu}

   \date{Received December 15, 1998; accepted }

   \maketitle

   \begin{abstract}

When does a GRB stop and its afterglow begin?  A GRB may be defined as
emission by internal shocks and its afterglow as emission by an external
shock, but it is necessary to distinguish them observationally.  With these
definitions irregularly varying emission (at any frequency) must be the GRB,
but smoothly varying intensity is usually afterglow.  The GRB itself and its
afterglow may overlap in time and in frequency, and distinguishing them
will, in general, require detailed modeling.

   \end{abstract}

%

At first glance there appears to be little difficulty in distinguishing
between GRB and their afterglows.  GRB were discovered in 1972, and are
observed in gamma-rays with detectors most sensitive to photons in the
range 100--1000 keV.  The observed durations of GRB range from milliseconds
to $\sim 1000$ s.  Afterglows were first observed in 1997 in the radio,
visible and X-ray bands, and have durations of hours to months.  These
appear to be very different phenomena, although causally
associated---afterglows follow GRB.

This phenomenological distinction between GRB and their afterglows is likely
to become insufficient.  It is based on the observations of GRB and of
afterglows by two distinct classes of instruments with distinct (and largely
non-overlapping) sensitivities:
\begin{enumerate}
\item GRB detectors have broad angular acceptance and little sensitivity
below 30 keV.  Their angular acceptance implies high
background levels.  This makes them comparatively insensitive to steady
sources of low flux, although very sensitive to transients of low fluence.
These properties are not failures of instrument design; rather, they
represent the optimal adaptation of detector technology to the observation
of unpredictable brief transients of high peak flux but low fluence
(compared to a steady source integrated over a long time).

\item Afterglows are detected with instruments which are sensitive to steady
sources of low flux and known position.  Observing such sources is the usual
problem in astronomy, and these are conventional astronomical instruments,
whose sensitivity depends on long temporal integration and high angular
resolution.  Their resolution discriminates against background.  They
cannot find unpredictable transients because their high
resolution limits their angular acceptance, but once steered to an afterglow
by a GRB detector they are sensitive to sources of low flux but long
duration (and therefore of comparatively large fluence).
\end{enumerate}

This instrumental distinction between GRB and their afterglows will
not survive when the gap between the two classes of instruments is bridged.
To some extent, this has already happened, as the X-ray detectors on
BeppoSAX are able to detect X-ray emission during the brief gamma-ray
duration of bright GRB within their field of view.  The X-ray and soft
gamma-ray bands overlap around 10--30 keV (the distinction between them is
largely a matter of definition), and X-ray emission during the detected
gamma-ray emission is generally considered to be just the low frequency
extrapolation of the gamma-ray emission (although there is a lively
controversy concerning the validity and interpretation of this
extrapolation; Katz \cite{K94a}, Preece, {\it et al.} \cite{P96}, Cohen,
{\it et al.} \cite{C97}, Preece, {\it et al.} \cite{P98}).

The detection of {\it simultaneous} visible counterparts has long been the
holy grail of GRB research.  Instruments are rapidly improving in
sensitivity and speed of response, and the prospect of rapid ($< 10$ s)
dissemination of accurate GRB coordinates (for example, from HETE-2) makes
it likely that they will be observed soon.

The phenomenological distinction between GRB and their afterglows will
disappear when the temporal gap between simultaneous and delayed (by hours
to days) observations is bridged.  Once the simultaneous radiation of a GRB
is detected, continuing to stare at that point on the sky will produce a
continuous intensity history interrupted only by dawn, bad weather or Earth
occultation.

Will there then be any possibility of distinguishing between GRB and their
afterglows, or any purpose in doing so?  I wish to argue that the answer to
both these questions in yes, provided the terms GRB and afterglow are
redefined as indicating the physical processes which produce the radiation.

There is no doubt that GRB involve relativistic outflow from a central
source of energy, and that the observed radiation is produced in optically
thin (except at radio frequencies) regions far from the central source.  In
order to tap the kinetic energy of relativistic matter it must exchange
momentum with some other matter or radiation, either nearly at rest in a
local observer's frame or also moving relativistically.  The former case is
called an external shock, and the matter at rest is generally assumed to be
either the surrounding interstellar medium or a non-relativistic outflow
produced by the GRB's progenitor.  The latter case is called an internal
shock, and the interaction is assumed to be between different portions of
the relativistic outflow, produced at different times and with differing
Lorentz factors.  Although the term ``shock'' is generally used, it is
neither necessary nor demonstrated that a shock occurs; streams of low
density matter may interpenetrate, exchanging momentum more gradually as a
consequence of plasma instability (such an instability is also required for
a shock, which must be collisionless because of the low densities).

In at least some GRB most of the early gamma-ray emission is produced by
internal shocks.  These GRB consist of several sharp and clearly separated
subpulses, often with intensity dropping to background levels between the
subpulses.  Fenimore, Madras and Nayakchin \cite{F96} and Sari and Piran
\cite{SP97} showed that such temporal behavior cannot be produced by an
external shock of plausible efficiency, no matter how clumpy the external
medium, thus refuting the original argument (Rees and M\'esz\'aros
\cite{RM92}, Katz \cite{K94b}) for external shock models, that interaction
with a heterogeneous medium can explain how a single class of similar
collapse or coalescence events could produce the observed ``zoo'' of
diverse GRB pulse profiles.  The radiation of these multipeaked GRB can only
be explained by internal shock (Rees and M\'esz\'aros \cite{RM94}) models,
in which the variability is attributed to modulation of a longer-lived
source of energy (Katz \cite{K97}).  Additional evidence for the internal
shock origin of multi-peaked GRB was presented at this meeting by Fenimore
and Ramirez-Ruiz (\cite{F99}) and by Quilligan, {\it et al.} (\cite{Q99})
who found that the properties of their subpulses do not evolve through a
burst, suggesting that subpulses are independent events, in effect
individual GRB.  This is inconsistent with external shock models, in which
the characteristic radii and time scales monotonically increase.

Internal shocks are rather inefficient ($\sim 20$\% for typical parameters)
in dissipating (a precondition for radiating) the kinetic energy of
relativistic outflows.  Because no GRB occurs in perfect vacuum, even if it
were to occur in an intergalactic medium, there is always matter for
external interaction.  The efficiency of radiation by this external
interaction may be low, particularly if the ambient density is low.  It is
natural to associate this external interaction (or shock) with the
phenomenologically defined afterglow because its duration is predicted to be
much longer than that of internal shocks, and because the smooth
single-peaked behavior of afterglows observed (to date) is consistent with
that predicted (Katz \cite{K94a,K94b}) for external shocks.

I suggest that it is useful to {\it define} GRB as the radiation produced by
internal shocks and afterglows as the radiation produced by external shocks.
Then, instead of arguing about nomenclature, we can engage in a more
scientifically productive discussion about the physical origin of the
various components of the observed radiation.  How can they be
distinguished?

A spiky temporal profile is an unambiguous indicator of an internal shock,
but that rule does not answer all questions.  There are GRB with smooth
single-peaked profiles, which can be explained by either internal or
external shocks,  The observed duration of some GRB is $\sim 1$ hour, and
others may last a day or longer (Katz \cite{K97}), so that duration is also
not sufficient to distinguish GRB from their afterglows.  In some bursts,
the internal and external shock emission may overlap in both spectrum and
in time.

Unfortunately, there appears to be no general rule for distinguishing
external from internal shock emission.  The predicted
asymptotic spectra with and without radiative cooling (Cohen, {\it et al.}
\cite{C97}) do not
distinguish between internal and external shocks.  Making this distinction
will require detailed spectral and temporal modeling to associate different
spectral components which are produced by the same physical process.
Identifying the physical origin of each will require hydrodynamic modeling
of the source.  This is likely to be a formidable task.

\begin{acknowledgements}
      This work was supported in part by NSF AST 94-16904.
\end{acknowledgements}

\end{document}